\documentclass[
 aip,
 apl,
 amsmath,amssymb,
 reprint,%
]{revtex4-1}

\usepackage{graphicx}
\usepackage{dcolumn}
\usepackage{bm}

\usepackage[utf8]{inputenc}
\usepackage[T1]{fontenc}
\usepackage{mathptmx}
\usepackage{etoolbox}
\usepackage{xr-hyper}
\usepackage{physics}
\usepackage{hyperref}
\usepackage{xcolor}
\usepackage{academicons}
\usepackage{epsfig}
\usepackage[outdir=./]{epstopdf}
\definecolor{orcidlogocol}{HTML}{A6CE39}
\newcommand{\orcid}[1]{\href{https://orcid.org/#1}{\textcolor[HTML]{A6CE39}{\aiOrcid}}}

\usepackage{lipsum} 

\usepackage{tikz,xcolor,hyperref}

\definecolor{lime}{HTML}{A6CE39}
\DeclareRobustCommand{\orcidicon}{%
	\begin{tikzpicture}
	\draw[lime, fill=lime] (0,0) 
	circle [radius=0.16] 
	node[white] {{\fontfamily{qag}\selectfont \tiny ID}};
	\draw[white, fill=white] (-0.0625,0.095) 
	circle [radius=0.007];
	\end{tikzpicture}
	\hspace{-2mm}
}

\foreach \x in {A, ..., Z}{%
	\expandafter\xdef\csname orcid\x\endcsname{\noexpand\href{https://orcid.org/\csname orcidauthor\x\endcsname}{\noexpand\orcidicon}}
}

\makeatletter
\def\@email#1#2{%
 \endgroup
 \patchcmd{\titleblock@produce}
  {\frontmatter@RRAPformat}
  {\frontmatter@RRAPformat{\produce@RRAP{*#1\href{mailto:#2}{#2}}}\frontmatter@RRAPformat}
  {}{}
}%
\makeatother
\begin{document}


\preprint{AIP/123-QED}

\title{Large Inverse Transient Phase Response of Titanium-nitride-based Microwave Kinetic Inductance Detectors}

\author{Jie Hu\orcidA{}}

\affiliation{GEPI, Observatoire de Paris, PSL Université, CNRS, 75014 Paris, France} 
\affiliation{Université de Paris, CNRS, Astroparticule et Cosmologie, F-75013 Paris, France}

\author{Faouzi Boussaha\orcidD{}}
 
\affiliation{GEPI, Observatoire de Paris, PSL Université, CNRS, 75014 Paris, France} 

\author{Jean-Marc Martin\orcidB{}}%
\affiliation{GEPI, Observatoire de Paris, PSL Université, CNRS, 75014 Paris, France} 

\author{Paul Nicaise\orcidC{}}
\affiliation{GEPI, Observatoire de Paris, PSL Université, CNRS, 75014 Paris, France} 

\author{Christine Chaumont}
\affiliation{GEPI, Observatoire de Paris, PSL Université, CNRS, 75014 Paris, France}

\author{Samir Beldi}
\affiliation{ESME Research Lab, 38 rue Molière, 94200 Ivry-sur-Seine - France}

\author{Michel Piat}
\affiliation{Université de Paris, CNRS, Astroparticule et Cosmologie, F-75013 Paris, France}

\author{Piercarlo Bonifacio}
\affiliation{GEPI, Observatoire de Paris, PSL Université, CNRS, 75014 Paris, France}

 \email[Authors to whom correspondence should be addressed:]{jie.hu@obspm.fr; faouzi.Boussaha@obspm.fr}

\date{\today}
\begin{abstract}
Following optical pulses ($\lambda=405~\text{nm}$) on titanium nitride (TiN) Microwave Kinetic Inductance Detectors (MKIDs) cooled down at temperatures $T \le T_c / 20$ ($T_c \simeq 4.6~\text{K}$), we observe a large phase-response highlighting two different modes simultaneously that are nevertheless related. The first corresponds to the well-known transition of cooper-pair breaking into quasi-particles which produces a known phase response. This is immediately followed by a large inverse response lasting several hundreds of microseconds to several milliseconds depending on the temperature. We propose to model this inverse pulse as the thermal perturbation of the superconductor and interaction with two level system (TLS) that reduces the dielectric constant which in turns modify the capacitance and therefore the resonance frequency. The ratio of the TLS responding to the illumination is on the order of that of the area of the inductor to the whole resonator
\end{abstract}

\maketitle 

Since its invention in 2003\cite{Day2003}, in order to reach ultimate performances, MKIDs are the subject of intensive studies and developments both at the theoretical and engineering levels\cite{Leduc, Zmuidzinas2012,Nika2020}.
MKIDs have particular interests on the optical and near-infrared band due to their fast response, low noise, intrinsic energy resolution and their scalability to large arrays\cite{Mazin2013,Gao2012,Beldi}.
A MKID is a superconducting LC resonator, usually consisting of a single superconducting layer deposited on a high resistivity silicon or sapphire substrate.
The operation principle of MKIDs is well-established in literature\cite{Zmuidzinas2012}: their resonance frequencies reduce after absorption of photons with an energy larger than twice the superconductor energy gap $\Delta=1.75k_BT_c$ through the creation of quasi-particles after breaking Cooper-pairs. This causes the frequency resonance $f_\text{r}  = 1/2\pi\sqrt{(L_g+L_k)C}$ to shift to lower frequencies, through the increase of its kinetic inductance $L_k$. $L_g$ is the geometric inductance.


However, for a MKID with two-level system (TLS)\cite{Gao2008} between the superconductor and the dielectric substrate, anomalous response has been reported\cite{Gao20081,Barends2008,Wang2013} in which the resonance frequency of the MKID increases when rising bath temperature $T_\text{bath}$ or under constant optical illumination.
In this paper, we report the anomalous response investigation of TiN MKIDs to short light pulses ($50-200~\text{ns}$ with $\lambda=405~\text{nm}$) at temperatures $T\le T_c/20$. The temperatures range in which the anomalous responses are observed is far below the usual MKIDs operating temperature ($\simeq T_c/10$) found in literature.
The anomalous response is composed of a normal phase response due to quasi-particle breaking and an inverse phase response, both observed simultaneously.
The inverse phase response we observed is considered to be related to thermal perturbation in the superconductor following the illumination, which excites the inverse phase response attributed to TLS.

\begin{figure}
    \includegraphics[width=\columnwidth]{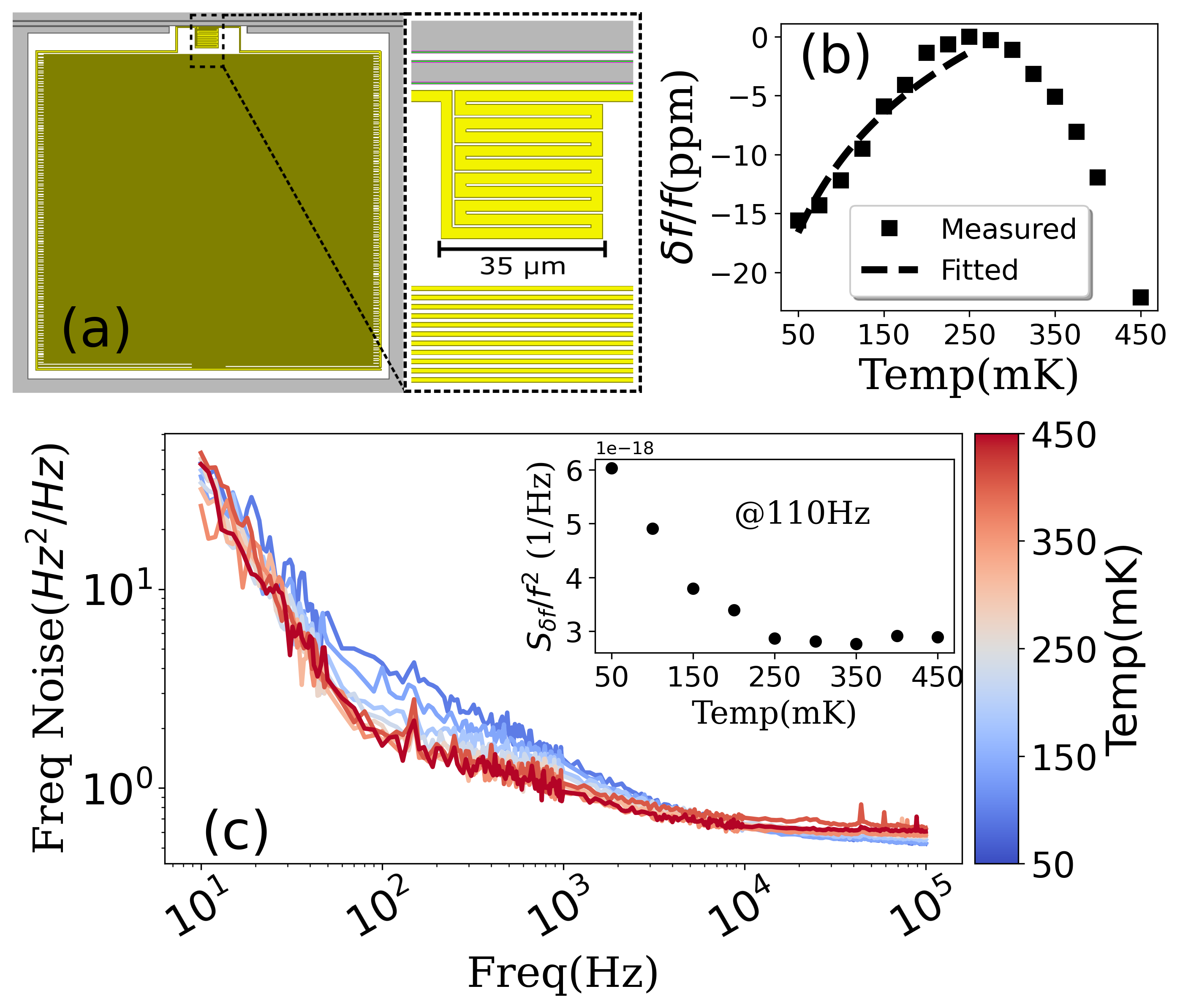}
    \caption{\label{fig:model} (a) Sketch of the investigated MKID. The resonator size is $523\times526~\mu\text{m}^2$. The sheet inductance of the film is about $7~\text{pH}/\square$. The interdigitated capacitor is formed of $1~ \mu\text{m}$ strips spaced by $1~ \mu\text{m}$ gap, the total capacitance is about $1.37~\text{pF}$. (b) Measured resonance fractional frequency shift versus temperature of a MKID with  $f_r$=$0.787~\text{GHz}$, $Q_l = 5000$ and $Q_i = 35000$ ($1/Q_l=1/Q_i + 1/Q_c$, $Q_c$ is the coupling quality factor to the feedline. The dashed line is the fit to the TLS giving by Eq. \ref{eq:TLS}, up to $250~\text{mK}$. This gives an estimated F$\delta$ of $3.58\times10^{-5}$. (c) Phase Noise Spectrum of the MKID in the dark. For frequency less than $50~\text{Hz}$, the phase noise is dominated by $1/f$ noise. TLS noise is observed in the frequency band between $50~\text{Hz}$ and $10~\text{kHz}$. The insert presents the noise evolution as a function of temperature at  $110~\text{Hz}$.}
\end{figure}

As shown in FIG.\ref{fig:model}-a, our MKIDs are made of a $35\times 32~\mu\text{m}^2$ meander connected in parallel to an interdigitated capacitor whose strip lengths are adjusted to a target resonance frequencies within the $0.7-1.0~\text{GHz}$ range.
The width of the meander line is $2.5~\mu\text{m}$ and its spacing is $0.5~\mu\text{m}$.
Both the meander and the capacitor are made of a $60~\text{nm}$ thick TiN layer deposited on $300~\mu \text{m}$ thick sapphire substrate, they are readout by a $50~\Omega$ coplanar transmission line made of $100~\text{nm}$ thick niobium layer. More details about physical parameters and fabrication process can be found in the paper by S. Beldi\cite{Beldi}.

The 20-pixel array is mounted in a gold-plated copper box and Al-wire-bonded to two $50~\Omega$ SMA connectors, and then cooled down to millikelvin in an adiabatic demagnetization fridge (ADR)\cite{Hu}. The stray magnetic field in the ADR is shielded by a pure niobium cylinder with a thickness of $1.5~\text{mm}$ around the chip. We first characterized the resonance frequency shift and noise versus bath temperature to assess the TLS effect within a MKID resonating at $f_r=0.787~\text{GHz}$. As shown in FIG.\ref{fig:model}-b, the resonance frequency increases as the bath temperature increases from 50 mK to 250 mK (i.e. $< T_c/20$), and then decreases for higher temperatures. In the first dynamics, the frequency shifts to higher frequencies because of TLS, while in the second, it moves to lower frequencies as the kinetic inductance increases. The change of the resonance frequency of the MKID due to TLS in the steady state can be well fitted using\cite{Gao2008}:
\begin{equation}\label{eq:TLS}
\frac{f_r(T)-f_r^{max}}{f_r^{max}} = \frac{F\delta}{\pi}\qty[\mathrm{Re}\Psi\qty(\frac{1}{2}+\frac{1}{2\pi i}\frac{\hbar\omega}{k_BT}) - \ln(\frac{\hbar\omega}{k_BT})],
\end{equation}
where $T$ is the temperature of the superconductor, $f_r(T)$ is the resonance frequency at $T$, $f_r^{max} = max(f_r(T))$, $F$ is the filling factor of the TLS and $\delta$ is the TLS-induced dielectric loss tangent at $T=0$  for weak non-saturating fields, $\Psi$ is the complex digamma function and $\omega=2\pi f_r(T)$ is the angular frequency of the input signal. The measured phase noise spectrum is shown in FIG.\ref{fig:model}-c. For frequency lower than $50~\text{Hz}$, the phase noise is dominated by $1/f$ noise. The TLS noise effect is particularly observed in the frequency band between $50~\text{Hz}$ and $10~\text{kHz}$. The insert shows how the noise evolves as a function of temperature, for example, at 110 \text{Hz}. The noise comparable with that in the literature\cite{Noroozian2012,Bueno} in dark. It clearly reduces as the temperature rises from $50~\text{mK}$ to $250~\text{mK}$ which is in agreement with the fractional frequency shift measurement.

\begin{figure*}
    \includegraphics[width=\textwidth]{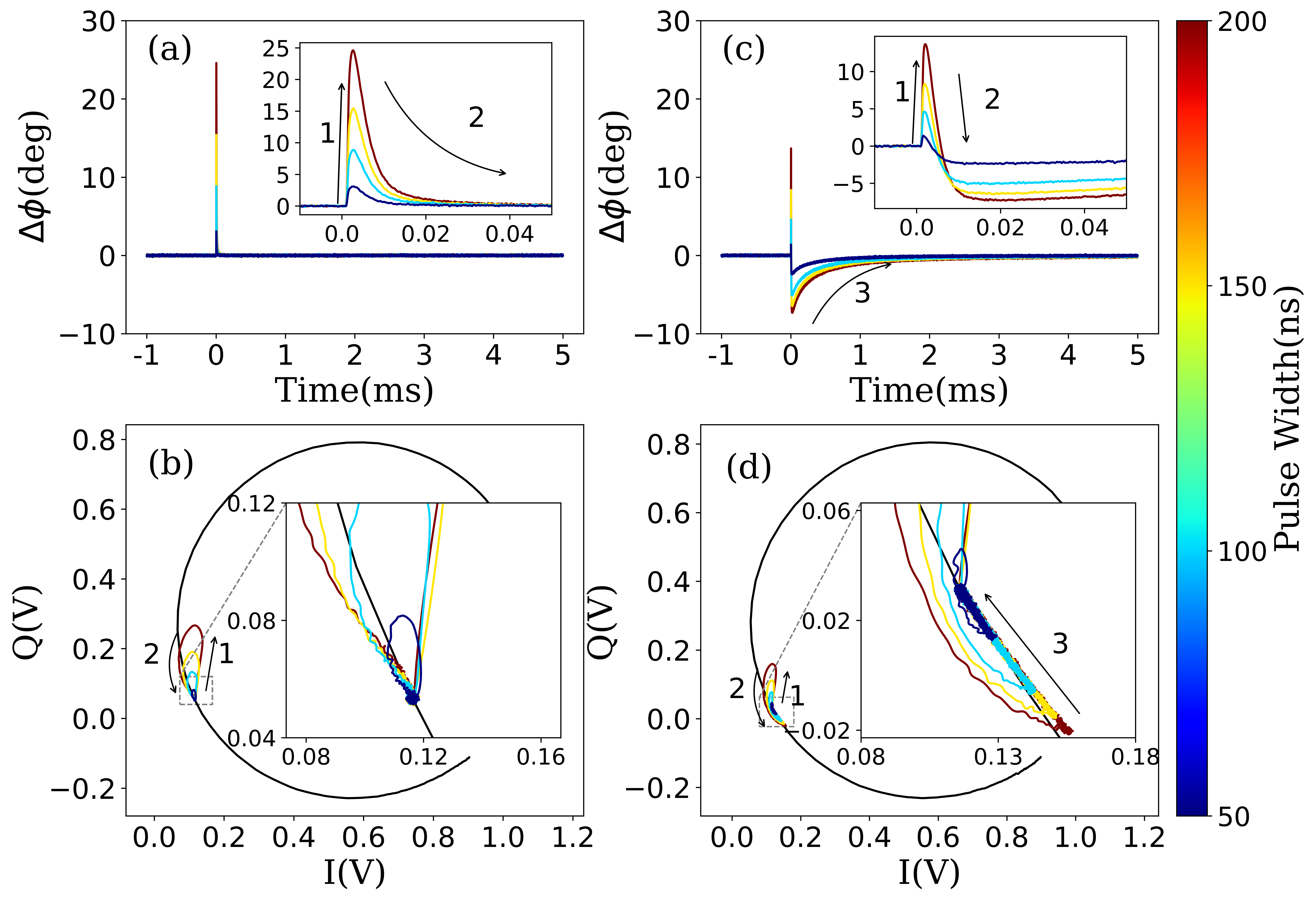} 
    \caption{\label{fig:phaseresponse} Measured averaged phase response of the resonator at $0.787~\text{GHz}$ and its trajectory on the IQ plane at $T_{bath} = 250~\text{mK}$ and $50~\text{mK}$. (a) Measured $\Delta \phi$ at $T_{bath} = 250~\text{mK}$. (b) Pulse trajectory at $250~\text{mK}$ in the IQ plane. (c) Measured $\Delta \phi$ at $T_{bath} = 50~\text{mK}$. (d) Pulse trajectory at $T_{bath} = 50~\text{mK}$ in the IQ plane.}
\end{figure*}

Using the homodyne routine scheme\cite{Mazin2005}, the array is illuminated by a $405~\text{nm}$ laser diode through an optical fiber placed about $35~\text{mm}$ above. The optical pulse is generated by modulating the diode at $61~\text{Hz}$. At room temperature, for a $100~\text{ns}$ pulse-width, the optical power $P_{opt}$ is estimated to be around $4~\text{pW}$ at the fiber output. The readout tone is placed at the minimum of the $S_{21}$ and its power ranges from $-110~\text{dBm}$ to $-85~\text{dBm}$. The nominal readout power is $P_r = -90~\text{dBm}$. The frequency of the pulse $61~\text{Hz}$ and the pulse is sampled by an oscilloscope at $100~\text{MHz}$. For $T_{bath} \geq 250~\text{mK}$, we always observed the well-known and standard response when MKIDs are exposed to pulse illuminations\cite{Day2003,Gao2012}.  FIG.\ref{fig:phaseresponse}-a and b show, respectively, the phase response and its trajectory in the IQ plane with bath temperature $T_{bath} \simeq 250~\text{mK}$ ($\simeq T_c/20$). The time constant of the rise edge $\tau_{res}$ is in the order of $1~\mu\text{s}$ and is determined by the response time of the resonator\cite{Mazin2005} which is $\tau_\text{res} = Q_l / \pi f_0 $. The maximum quasiparticle lifetime $\tau_\text{qp}$ is around 10 µs which is in good agreement with values reported elsewhere \cite{Leduc,Beldi}.  
In the IQ plane, this response is reflected in the phase change $\Delta \phi$ along with a change in the radius of the resonance circle \cite{Gao2008} $r = |Q_l/2Q_c|$ due to dissipation caused by the quasi-particles. Furthermore, this standard phase change which originates from Cooper pairs breaking\cite{Day2003} indicates that the frequency resonance shift towards lower frequencies.

\begin{figure}
    \includegraphics[width=\columnwidth]{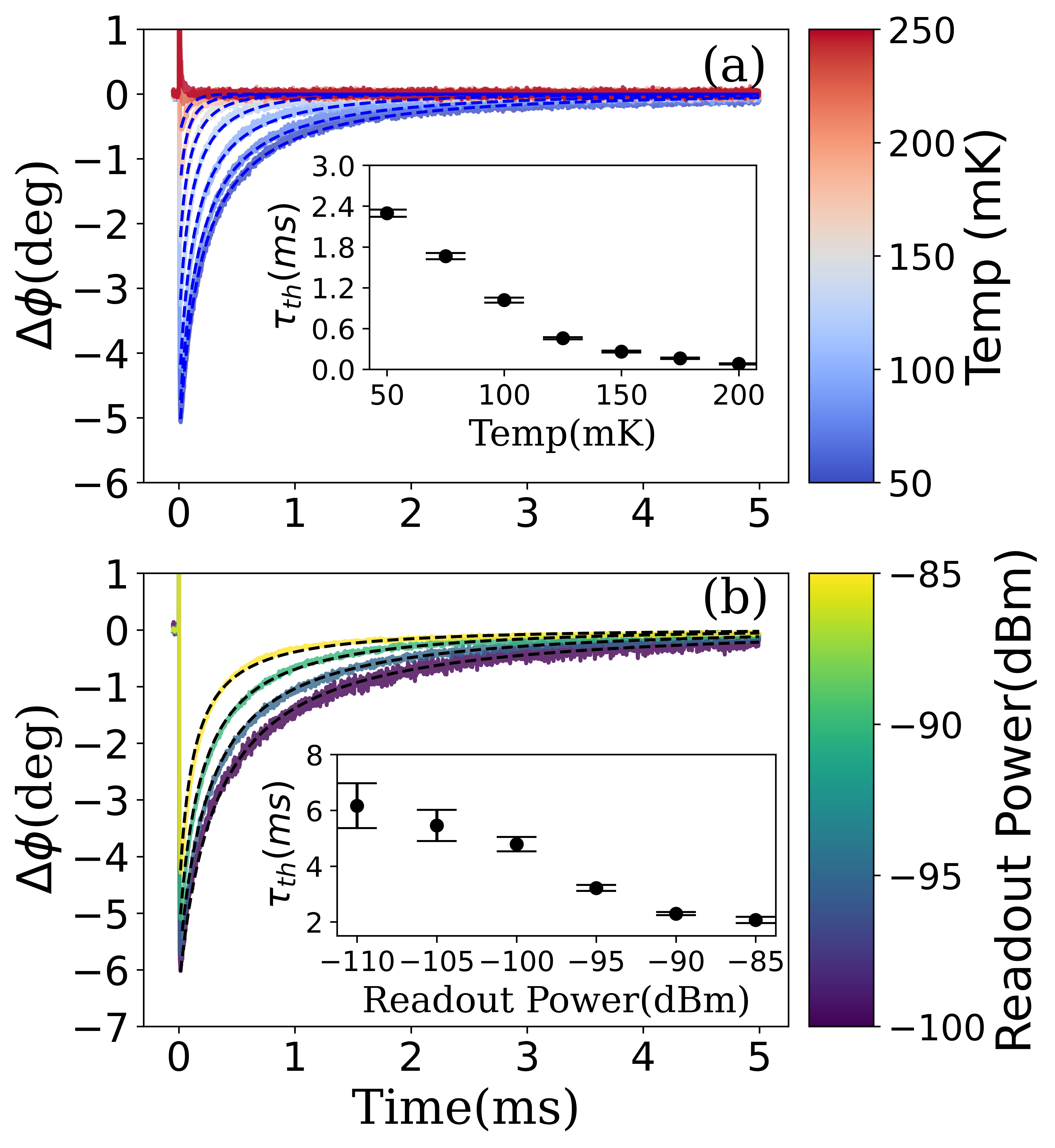} \\
    
    \caption{\label{fig:inversepulsevspowerandtemp} Phase response of the MKID at different readout power and bath temperatures. The dashed lines are the fitted results with equation \eqref{eq:dphi}. The insect is the fitted time constant $\tau_\text{th}$ at different readout power and temperatures. The optical power at the fiber output is estimated to be $4~\text{pW}$. (a) Phase response of the MKIDs from $50~\text{mK}$ to $250~\text{mK}$ with $P_r = -90$ dBm.(b) Phase response of the MKID with readout power range from $-110~\text{dBm}$ to $-85~\text{dBm}$ at $50~\text{mK}$.}
\end{figure}

With the same readout power $P_r$ and illumination, the phase response and trajectory in the IQ plane of the resonator at $T_{bath}=50~\text{mK}$ is shown in FIG.~\ref{fig:phaseresponse}-c and d, respectively. We notice that the standard response extends and gives rise to a second inverse phase response of the same order in magnitude.
The time constant of the rise edge of the inverse phase response is about $3~\mu\text{s}$ , much longer than $\tau_\text{res}$.
The fitted relaxation time constant of the inverse pulse is around $2.3~\text{ms}$. The inverse response indicates that the resonance frequency moves in the opposite way towards higher frequencies, which is the typical response of TLS at the steady state as presented above. In the IQ plane, its trajectory moves almost along with the resonance circle, which means that the quality factor of the resonator almost remains constant. It is therefore probably not related to the Cooper-pairs breaking in the superconductor. FIG.~\ref{fig:inversepulsevspowerandtemp} presents the inverse response at different temperature and readout powers. It shows a strong correlation between $\Delta \phi$ magnitude and the TLS strength highlighted by the fractional frequency shift measurements versus temperature shown in FIG.\ref{fig:model}-b. On the other hand, measurements as a function of readout power show that the higher the power, the lower $\Delta \phi$ (i.e. the lower TLS effect) which is also consistent with results reported elsewhere\cite{Gao2008}.



The inverse phase response is probably due to phonon-TLS interaction\cite{Li2016} where TLS are excited by phonons generated by the absorption of photons into the superconductor bulk, emitted when quasiparticles recombine into Copper pairs\cite{BCS}, which is possible as it is indicated by the succession of the two stages of the phase response. As expressed by Eq.~\ref{eq:TLS}, TLS affect the substrate dielectric constant $\epsilon_r$ which in turn affects the capacitance and therefore $f_r$. 

The inverse pulse is therefore almost dissipationless and is predominantly caused by the capacitor change. However, the little deviation observed at the beginning of the IQ trajectory indicates that the quality factor of the MKID is lower than that in the steady state, which is possibly due to the superconductor are not thermalized and there still are thermally excited quasiparticles\cite{Zmuidzinas2012}.
The inverse response can be modeled as a thermal perturbation of the superconductor that leads to the excitation of TLS in the resonator. The thermal power flow into the superconductor is\cite{Wandui2020} 
\begin{equation}\label{eq:power}
C_S\dv{T}{t} = -K\qty(T^n-T_\text{bath}^n) + P_s + P_\text{opt},
\end{equation}
where $K$ is geometry-dependent factor, $n$ the power-law coefficient  ($4\leq n\leq 6$)\cite{Irwin1995}, $C_s$ is the heat capacity of the superconductor, $P_s=2P_rQ_l^2/Q_iQ_c$ is the dissipated power in the MKID\cite{Visser2014} and $P_\text{opt}$ is the illuminated optical power.
The heat capacity $C_S$ consists of that of electrons whose contribution is very negligible,  as well as phonons and TLS at the interface of the superconductor and the dielectric. Here we assume the temperature of the TLS is the same as that of the phonons. Detailed analysis of the heat capacity can be found in Sec. A in the supplementary material.  In the steady state, the temperature of the TiN is $T_s = (P_s/K + T_{bath}^n)^{1/n}$. 
We focus our model of the anomalous phase response for times $t \geq t_0$ where we can consider that the normal MKID response to photons is over. For a small thermal perturbation in TiN caused by phonons in the form of a time-dependent temperature increase of $T_1$, we set $T({t}) = T_s + T_1({t})$. With Eq.~\eqref{eq:power}, $T_1({t})$ can be written as:
\begin{equation}\label{eq:dT1dt}
\dv{T_1}{t} \approx -\frac{G}{C_S}\qty(T_1 + \frac{(n-1)}{2T_s}T_1^2),
\end{equation}
where $G = nKT_s^{n-1}$ is the thermal conductance. 
Using $\Delta f = -\Delta\phi f_r(T_s)/4Q$\cite{Mazin2005} and the analytical solution of Eq.~\eqref{eq:dT1dt}, we can then write $\Delta\phi(T_s,t \geq t_0)$ to fit the data in FIG.~\ref{fig:inversepulsevspowerandtemp} as:
\begin{equation}\label{eq:dphi}
\eval{\Delta\phi(T_s,t)}_{t \geq t_0} \approx -\frac{4Q_l}{f_0}\pdv{f_r}{T} \frac{T_0e^{-(t-t_0)/\tau_\text{th}}}{1+\frac{(n-1)T_0}{2T_s}\qty(1-e^{-(t-t_0)/\tau_\text{th}})},
\end{equation}
where $T_0=T_1(t=t_0)$ and $\tau_\text{th}=C_S/G$.
In Eq.~\ref{eq:dphi}, the time constant $\tau_\text{th}$ of the inverse pulse response is not directly related to the TLS but to the thermal flow between the superconductor and the substrate. Thus, the direction of the phase response is determined by the term $\partial f_r/\partial T$. The model also shows that the inverse pulse can only be observed when its time constant $\tau_{th}$ is larger than the quasiparticle recombination time $\tau_{qp}$. The full calculation of the model can be found in Sec.B in the supplementary material.\\
As shown in FIG.~\ref{fig:inversepulsevspowerandtemp}, this equation allows us to fit quite well the inverse pulse response. The fitted $\tau_\text{th}$ at $50~\text{mK}$ is around $2.3~\text{ms}$. This long time constant may originate from the relative long relaxation time constant of the TLS, the phonon mismatch between the TiN and the substrate\cite{Visser2021} or the fact that the exsistance of the TLS at the interface between the superconductor and substrate reduces the thermal conductance. $\tau_\text{th}$ then decreases significantly as the temperature bath rises approaching zero at around $225~\text{mK}$, before the typical TLS response completely vanish (around $250~\text{mK}$), as is shown in FIG.~\ref{fig:model}-b. 
Furthermore, the measured $\tau_\text{th}$ is in consistent with the noise shown in FIG.~\ref{fig:model}-c, as the time constant reduces to be less than $100~\mu\text{s}$ at $200~\text{mK}$, it is larger than $10~\text{kH}z$ in frequency. Thus, the TLS noise is rolled off by the resonator. FIG.~\ref{fig:inversepulsevspowerandtemp}-b shows $\tau_{th}$ decreases significantly as the readout power increases as the TLS are saturated, either reducing the heat capacity or improving the thermal conductance between the TiN and the sapphire. The long time constant of the TLS could be the reason of the excess low frequency noise of TiN MKIDs under illumination\cite{Bueno}.

The fitted $T_0$ ranges from $200~\text{mK}$ ($n=6$) to $350~\text{mK}$ ($n=4$), as is shown in FIG.~\ref{fig:cal_dFdT}-a. More detail can be found in Sec.C in supplementary material. The value of n needs to be further investigated. The ratio of the $\partial f_r /\partial T$ from the fitting parameter to the value obtained by changing the bath temperature, i.e. from FIG. \ref{fig:model}-b is shown in FIG.~\ref{fig:cal_dFdT}-b, which is on the same order to the ratio of area of the inductor to that of the whole resonator. The $\partial f_r /\partial T$ is larger as TLS in the whole resonator are responding to the increasing temperature. It strongly indicates that the inverse phase response are localized in the resonator, which is probably from the vicinity of the inductor. Thus, the fraction can be considered as the ratio of the area contributed to the inverse phase response. The ratio below 100 mK tends to be overestimated as the $\partial f_r / \partial T$ in the steady state is small while the temperature change is relatively large.

With this assumption, we estimate the thermal conductance $G\approx0.21 \ \text{pW}/\text{K} $ ($n=4$) and $G \approx 2.1 \ \text{fW/K}$ ($n=6$) at $50~\text{mK}$ by fitting the temperature dependence of thermal time constant and tends to decrease with increasing value of $n$.


\begin{figure}
    \centering
    \includegraphics[width = \columnwidth]{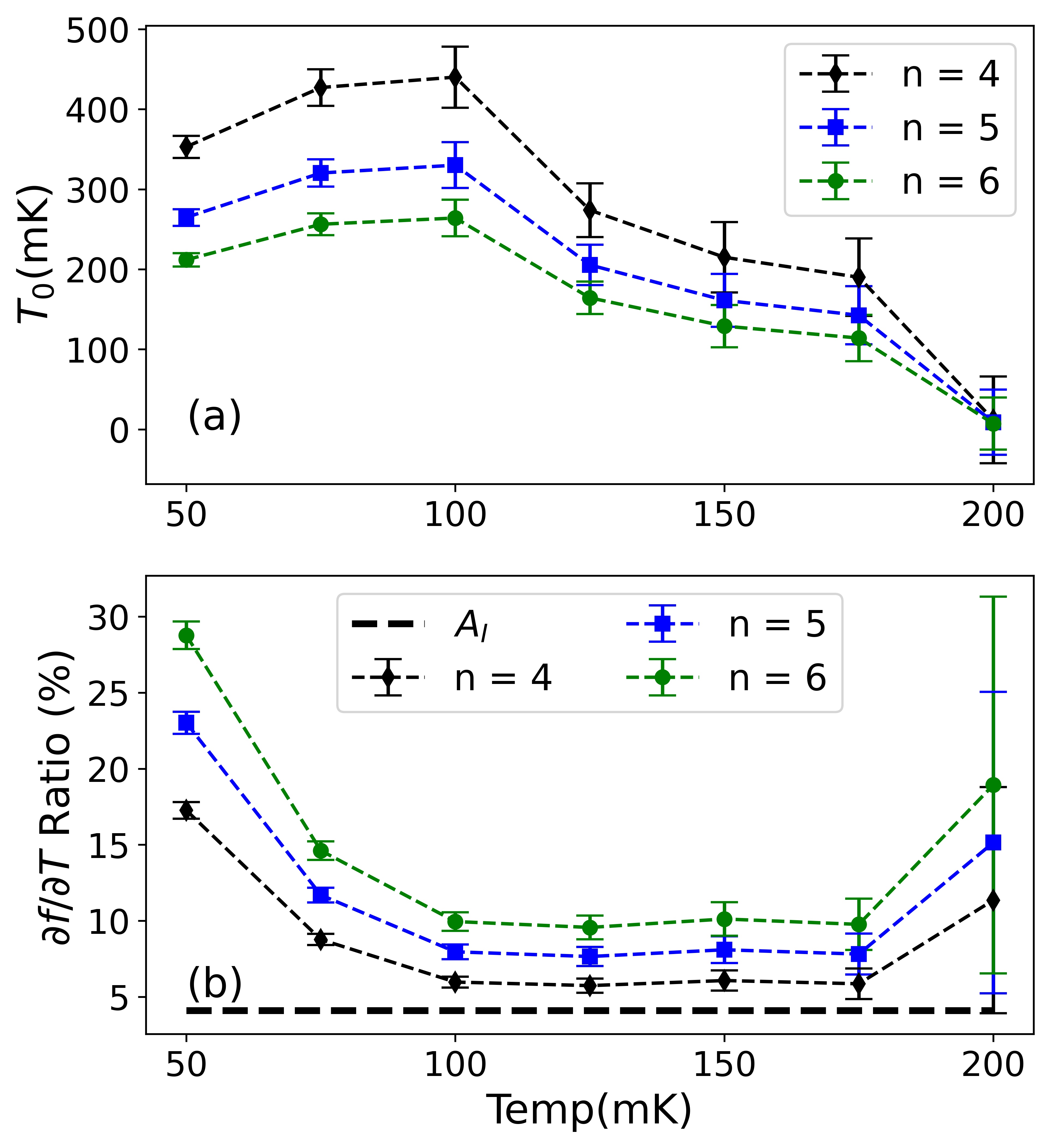}
    \caption{(a) Estimated $T_0$ from the fitted parameters. (b)Estimated fraction of $\partial f_r/\partial T$ to the frequeny shift from TLS by changing the bath temperature. $A_I \approx 4.1 \%$ is the area ratio of the inductor to the whole area of the resonator.}
    \label{fig:cal_dFdT}
\end{figure}

In conclusion, we have investigated optical pulse response of a MKID with stoichiometric TiN on a sapphire substrate from $50~\text{mK}$ to $450~\text{mK}$. The response from the quasi-particle breaking and the TLS response has been observed simultaneously with optical pulse illumination below $250~\text{mK}$.
The inverse pulse response from the TLS response is empirically modeled as thermal perturbation in the superconductor.
The fitted thermal constant of superconductor is on the order of $2~\text{ms}$ around $50~\text{mK}$ and decreases to zero around $250~\text{mK}$. The analysis also shows the heat capacity from the TLS becomes significant at low temperature. This can also serve as a basis for future investigations on phonon dynamics and phonon-TLS interactions within the superconductor/substrate interface. Supplementary studies of this TLS response of the detectors could lead to improve the performances of MKIDs.

\section*{supplementary material}

See supplementary material for the Sec. A, B, C. 


\begin{acknowledgments}
We would like to thank Damien Prele and Manuel Gonzalez for helpful discussion about cryogenic measurements, as well as Florent Reix, Josiane Firminy and Thibaut Vacelet for the assembly and mounting the devices. This work is supported by European Research Council (ERC) and UnivEarhS Labex program.

\end{acknowledgments}

\section*{Data Availability Statement}
The data that support the findings of this study are available from the corresponding authors upon reasonable request.
\section*{Reference}
\bibliography{aipsamp}

\setcounter{equation}{0}
\setcounter{figure}{0}

\renewcommand{\figurename}{\renewcommand{\thefigure}{FIS.S\arabic{figure}}}

\renewcommand{\tablename}{\renewcommand{\thetable}{TABLE.S\arabic{table}}}
\renewcommand{\theequation}{S\arabic{equation}}

\clearpage

\section*{Supplementary Material}
\subsection{Superconductor heat capacity analysis}\label{Heat capacity analysis}
The heat capacity of a superconductor with TLS consists of the heat capacity of the electron, the phonons and the TLS. For temperature $T<0.05T_c$, the heat capacity of the electron in the superconductor can be estimated as\cite{Tinkham2004}:
\begin{align}
    C_V^{el}=1.34 N_{el}\gamma T_c \qty(\frac{\Delta(0)}{k_B T})^{3/2} e^{-\frac{\Delta(0)}{k_B T}},
\end{align}
where $N_{el}$ is the number of the free electron per unit volume, $\gamma=\pi^2k_B^2/2E_F$ is the coefficient of the linear specific heat of a normal metal, $E_F$ is Fermi energy and $\Delta(0) = 1.75k_B T_c $ is the energy gap of the superconductor. We estimate $N_{el}=7.955\times 10^{10}~\mu\text{m}^{-3}$, $\gamma=0.603~\text{mJ}\cdot\text{mol}^{-1}\cdot\text{K}^{-2}$ with $E_F = 5.865~\text{eV}$\cite{Patsalas2015}. 

The heat capacity of the superconductor from phonons far below the Debye temperature $\theta_D \approx 485~\text{K}$) can be estimated as\cite{kittel1996}:
\begin{align}
    C_V^{ph}=234N_{ph}k_B \frac{T^3}{\theta_D^3},
\end{align}
where $N_{ph}$ is the number of the phonon mode per unit volume. We estimate the number of TiN phonon modes in our film to be $N_{ph}=1.553\times10^{11}\mu m^{-3}$.

\begin{figure}
    \centering
    \includegraphics[width=\columnwidth]{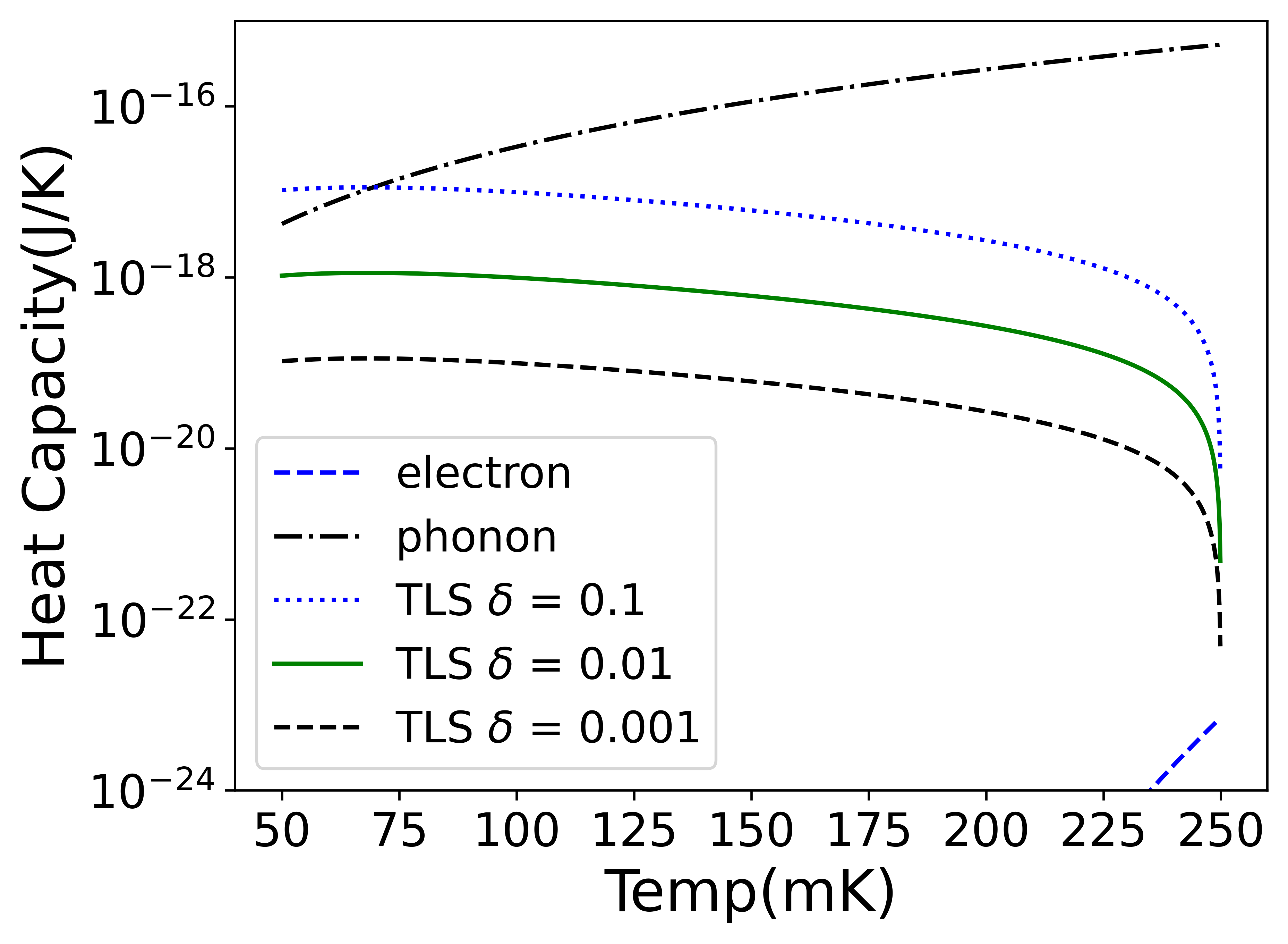}
    \caption{Estimated Heat Capacity for the phonons, TLS and the electron in the TiN including the capacitor and the inductor.}
    \label{fig:Heat capacity}
\end{figure}

The specific heat of $N$ particles in a two level system of energy difference $E$ between the two levels can be estimated as\cite{Phillips1987}:
\begin{align}
    C_{TLS} = k_BN\qty(\frac{E}{k_BT})^2e^{-E/k_BT}.
\end{align}
From the standard tunnelling model of TLS\cite{muller2019}, the loss tangent of the TLS loaded material can be estimated as:
\begin{align}
    \delta = \frac{\pi P d^2}{3\epsilon},
\end{align}
where the constant $P$ is the density of states of the TLS in the material, $d$ is the electrical dipole of the TLS and $\epsilon$ is the dielectrical constant of the material.  Then the heat capacity from the TLS can be estimated as :
\begin{align}\label{CV_TLS}
    \begin{aligned}
     C_V^{TLS} &= \int_{E_1}^{E_2}C_{TLS}P\mathrm{d}E, \\
               &= \frac{3\epsilon\delta}{\pi d^2}\int_{E_1}^{E_2}k_B\qty(\frac{E}{k_BT})^2e^{-E/k_BT}\mathrm{d}E.
    \end{aligned}
\end{align}
We assume the TLS temperature to be the same as the phonons temperature, the energy difference of the TLS is on the order of the thermal phonons, i.e. $E\approx k_BT$. Thus, we assume $E_1\approx k_BT$ with the same temperature of the superconductor. Then Eq.~\eqref{CV_TLS} can calculated as 
\begin{align}
    C_V^{TLS} = \frac{3\epsilon\delta k_B^2T}{\pi d^2}\qty{1.839 - \qty[\qty(\frac{E_2}{k_BT}+1)^2 + 1]e^{-E_2/k_BT}}
\end{align}
where $E_2\approx k_BT_2$ with $T_2\approx250 \ \text{mK}$ in our case. \\
We also assume electrical dipole as $d=ed_0$, e is the electron charge and $d_0$ is the lattice constant of TiN. 
The calculated heat capacity from the electron, the phonons and the TLS is shown in FIGS.~\ref{fig:Heat capacity}

The detailed parameters to estimate the heat capacity is shown in Table \ref{TiN parameters}. It can be seen that the heat capacity from electron can be neglected. It should also be noted that the heat capacity of the TLS is proportional to the loss tangent of the material. 

\begin{table}
\caption{Parameters of the TiN MKIDs }\label{TiN parameters}
\begin{center}
    \begin{tabular}{c c}
    
    \hline
    TiN lattice constant     &  $0.428 \ \text{nm}$  \\
    \hline
    TiN thickness & 60 nm \\
    \hline
    Meander area     &  $925~\mu\text{m}^2$ \\
    \hline
    Capacitor area    & $ 126959~\mu\text{m}^2$ \\
    \hline
    TiN Volume      &  $7667~\mu\text{m}^3$ \\
    \hline
    TiN Density     &  $5.319~\text{g/cm}^3$ \\
    \hline
    Fermi velocity  &  $1.34 \times 10^6~\text{m/s}$ \\
    \hline 
    TiN mole weight & $ 61.874~\text{g/mol}$ \\ 
    \hline

\end{tabular}
\end{center}
\end{table}

\subsection{Full calculations}\label{appendix}

The resonance frequency shift of the resonator is determined by the temperature of the superconductor and the number of quasiparticles in the superconductor. It can be written as:
\begin{align}
    \Delta \phi(t) & = -\frac{4Q_l}{f_0}\Delta f_r(t) \nonumber \\
                   & = -\frac{4Q_l}{f_0}\Delta f_r(T(t), N_{qp}(t)), 
\end{align}
where $\Delta f(T(t),N_{qp}(t)) = f_r(T(t),N_{qp}(t))-f_0$, $f_0=f_r(T_s,N_{qp}^0)$ is the resonance frequency when the superconductor in the dark, $N_{qp}(t)$ is the quasiparticle number in the superconductor and $N_{qp}^0 = N_{qp}(t = 0)$.

With a small thermal perturbation due to phonons and quasiparticles, i.e. $T(t)=T_s+T_1(t),~N_{qp} = N_{qp}^0 + \Delta N_{qp}(t)$, we can make a first order Taylor development over the temperature and the quasi-particle number as:
\begin{align} \label{eq:annexegeneral}
\Delta\phi \approx &-\qty(T-T_s)\times\eval{\pdv{(\Delta \phi)}{T}}_{T=T_s} \nonumber
\\
&-\Delta N_{qp} \times\eval{\pdv{(\Delta\phi)}{N_{qp}}}_{N_{qp} = N_{qp}^0} \nonumber
\\ 
\approx&  -\frac{4Q_l}{f_0}\qty(T_1\eval{\pdv{f_r}{T}}_{T=T_s} +\Delta N_{qp}\eval{\pdv{f_r}{N_{qp}}}_{N_{qp} = N_{qp}^0}).
\end{align}

The first term in Eq.~\eqref{eq:annexegeneral} corresponds to change of the capacitor due to the superconductor temperature change mainly due to phonons, whereas the second term corresponds to the change of the kinetic inductance caused by Cooper pair breaking. The two terms are dependent due to the quasiparticle and phonon interactions as well as the thermal exitation of the quasiparticles. The decay of the $\Delta N_{qp}$ is dominated by the quasiparticle lifetime $\tau_{qp}$.

However, if we are at a time $t_0$ when the number of the excess quasiparticle generated by the optical illumination is negligible, then $\Delta N_{qp}(t \geq t_0) \simeq 0$, therefore, Eq.~\eqref{eq:annexegeneral} becomes:
\begin{equation}
\label{eq:annexefinal}
\eval{\Delta\phi(T_s,t)}_{t \geq t_0} \approx -\frac{4Q_l}{f_0}\times\pdv{f_r}{T} T_1(t).
\end{equation}

Furthermore, for small thermal perturbation of Eq.~\eqref{eq:power}, can be expressed as:
\begin{align}\label{append:dT_expand}
C_S\frac{\mathrm{d}T_1}{\mathrm{d}t} =& -K[(T_s+T_1)^n - T_{bath}^n] + P_s + P_{opt} \nonumber \\
                                          =&  -K\qty(nT_s^{n-1}T_1 + \frac{n(n-1)}{2}T_s^{n-2} T_1^2+...) \nonumber \\
                                          &-K(T_s^n -T_{bath}^n) + P_s + P_{opt} \nonumber \\
                                           =& -nKT_s^{n-1}T_1 - \frac{n(n-1)K}{2}T_s^{n-2}T_1^2 + ...
\end{align}
where the term $-K(T_s^n-T_{bath}^n) + P_s = 0$ as that it is in the steady state and $P_{opt} = 0$ as there is no illumination. Then Eq.~\eqref{eq:dT1dt} can be obtained as:
\begin{equation}
\dv{T_1}{t} \approx \frac{G}{C_S}\qty(T_1 + \frac{(n-1)}{2T_s}T_1^2),
\end{equation}
Which lead to the analytical expression of $T_1(t)$ as:
\begin{equation}\label{eq:dphi_appendix}
T_1(t) = \frac{T_0e^{-(t-t_0)/\tau_{th}}}{1+\frac{(n-1)T_0}{2T_s}\qty(1-e^{-(t-t_0)/\tau_{th}})}.
\end{equation}
where $t_0$ is the initial time (where the approximations are valid), $T_0=T_1(t=t_0)$ and $\tau_\text{th}=C_S/G$.

Together with Eq~\eqref{eq:annexefinal} and Eq~\eqref{eq:dphi_appendix}, the Eq.~\eqref{eq:dphi} can be obtained, which we rewrite here as follows:
\begin{equation}
\eval{\Delta\phi(T_s,t)}_{t \geq t_0} \approx -\frac{4Q_l}{f_0}\pdv{f_r}{T}\frac{T_0e^{-(t-t_0)/\tau_\text{th}}}{1+\frac{(n-1)T_0}{2T_s}\qty(1-e^{-(t-t_0)/\tau_\text{th}})}.
\end{equation}
It can be seen that the sign of $\Delta \phi$ is determined by the term $\partial f_r/\partial T$, which means if there is TLS in the resonator, there would be inverse phase response. It should also be noted that the inverse phase response can be observed only when $\tau_{th}$ is larger than the quasiparticle recombination time $\tau_{qp}$.
This not only explains why the inverse response can be seen in temperature regime where the TLS is present, but also explains that the inverse response disappears even when there is still TLS observed around $225~\text{mK}$. 




\subsection{Fitting Parameters}\label{Fitting Parameters}

\begin{figure}
    \centering
    \includegraphics[width = \columnwidth]{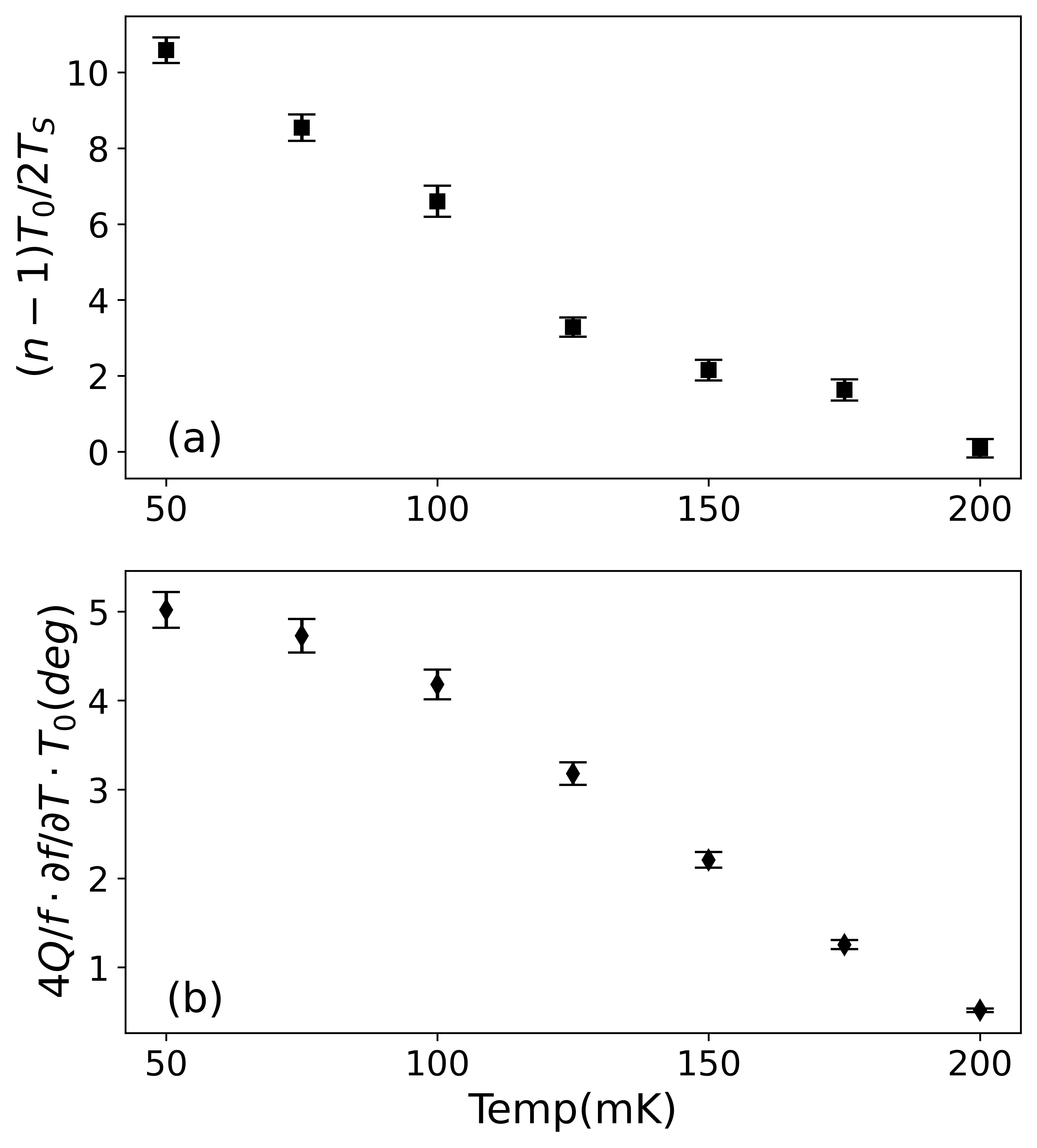}
    \caption{Detailed Fitting Parameters. (a) Fitted parameter $C=(n-1)T_0/2T_S$. (b) Fitted parameter $B=4Q/f \cdot \partial f_r/\partial T \cdot T_0$}
    \label{fig:fitting parameters}
\end{figure}

The fitted parameters are $\tau_\text{th}$, $B = 4Q/f_0\cdot \partial f_r /\partial T\cdot T_0 $ and $C = (n-1)T_0/2T_S$. They are shown in FIGS.~\ref{fig:fitting parameters}, the errorbar is from the standard least square fitting. It can be seen that as the temperature increases, the maximum inverse phase change decreases.

\begin{figure*}
    \centering
    \includegraphics[width = 6in]{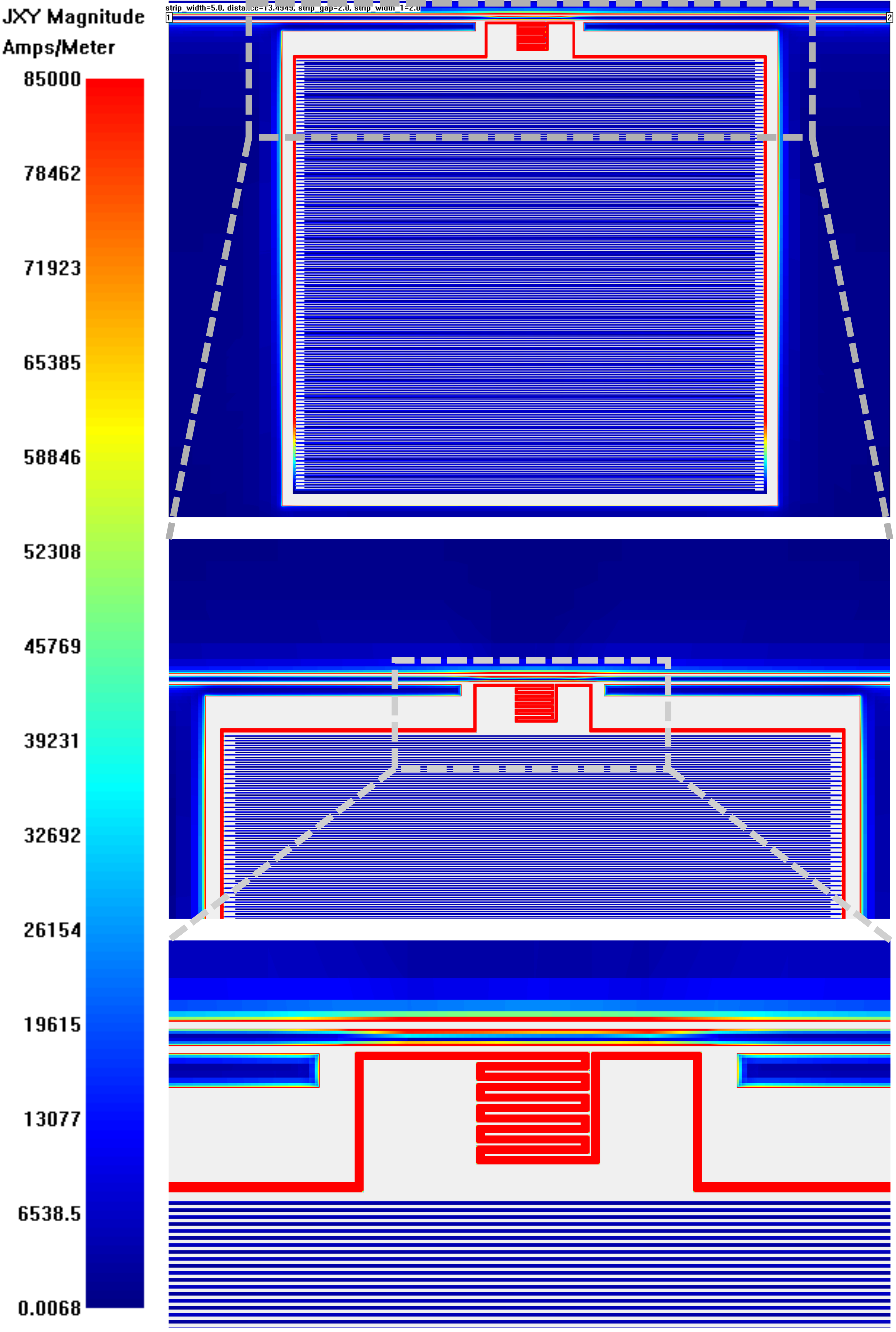}
    \caption{Current density of the resonator from SONNET}
    \label{fig:Current density}
\end{figure*}

We assume $T_S \approx T_{bath}$, $T_0 \approx 2C \cdot T_{bath}/(n-1)$ calculated with $n=4,5,6$ is shown in FIG.~\ref{fig:cal_dFdT}-a. It can seen that with smaller n, the estimated $T_0$ increases. After considering a value of n, we estimate $T_0$ from $C$. We calculate the value of $\partial f_r / \partial T$ following the light pulse to be different as in FIG.~\ref{fig:model}-b, we suppose this discrepancy due to the excitation of a lower quantity of TLS in the light pulse case. Thus, the $\partial f_r / \partial T$ obtained from the fitted parameters will be a fraction of the $\partial f_r / \partial T$ obtained in FIG.~\ref{fig:model}-b, as is shown in FIG. \ref{fig:cal_dFdT}b. It can be seen that the $\partial f / \partial T$ ratio is on the same order then the ratio of the inductor area to the whole resonator area. This is a strong indication that the inverse phase response is originated from the vicinity of the inductor. The inductor is considered as the superconductor with large current density, including the meander and its arms as you can see in FIGS.~\ref{fig:Current density} which shows the current density in the resonator from a SONNET simulation. The area of the inductor is about 4.1\% of the total area of the resonator, which is about 4 times more than the area of the meander. 

\end{document}